\documentclass[aps,preprint,superscriptaddress,showpacs,nofootinbib]{revtex4-1}
\usepackage{graphicx}
\usepackage{graphicx}
\usepackage{color}
\usepackage{mathtools} 
\usepackage{amsmath}
\usepackage{caption}
\usepackage{subcaption}
\usepackage{float}
\usepackage{subfig}
\usepackage{ulem}
\usepackage{color}
\definecolor{My_red}        {cmyk}{0.00,1.00,1.00,0.20}

\usepackage{pstricks}
\usepackage{bbm}

\newcommand{\bmat}{\left(\begin{array}}
\newcommand{\emat}{\end{array}\right)}
\newcommand{\beq}{\begin{equation}}
\newcommand{\eeq}{\end{equation}}

\newcommand{\VEV}[1]{\langle  #1 \rangle}
\newcommand{\mfrac}[2]{\frac{ \mbox{$#1$} }{ \mbox{$#2$} }}

\def\bwt{\begin{widetext}}
\def\ewt{\end{widetext}}
\def\be{\begin{equation}}
\def\ee{\end{equation}}
\def\bea{\begin{eqnarray}}
\def\eea{\end{eqnarray}}
\def\bean{\begin{eqnarray*}}
\def\eean{\end{eqnarray*}}
\def\bary{\begin{array}}
\def\eary{\end{array}}
\def\bit{\begin{itemize}}
\def\eit{\end{itemize}}

\def\su5u1{SU(5) \times U(1)}
\def\fsu5u1{SU(5) \times U(1)'}
\def\so10{SO(10)}
\def\sq20{SO(10) \times SO(10)}

\def\nn{\nonumber}

\def\bwt{\begin{widetext}}
\def\ewt{\end{widetext}}
\def\be{\begin{equation}}
\def\ee{\end{equation}}
\def\bea{\begin{eqnarray}}
\def\eea{\end{eqnarray}}
\def\bean{\begin{eqnarray*}}
\def\eean{\end{eqnarray*}}
\def\bary{\begin{array}}
\def\eary{\end{array}}
\def\bit{\begin{itemize}}
\def\eit{\end{itemize}}

\def\su5u1{SU(5) \times U(1)}
\def\fsu5u1{SU(5) \times U(1)'}
\def\so10{SO(10)}
\def\sq20{SO(10) \times SO(10)}

\usepackage{amssymb}

\begin{document}

\title{The Diboson Excesses in an Anomaly Free Leptophobic Left-Right Model}

\author{Kasinath Das}

\affiliation{Harish-Chandra Research Institute and Regional Centre for
Accelerator-based Particle Physics, Chhatnag Road, Jhusi, Allahabad 211019, India}

\author{Tianjun Li}

\affiliation{State Key Laboratory of Theoretical Physics and 
Kavli Institute for Theoretical Physics China (KITPC),
Institute of Theoretical Physics, Chinese Academy of Sciences, 
Beijing 100190, P. R. China}

\affiliation{
School of Physical Electronics, University of Electronic Science and Technology of China, 
Chengdu 610054, P. R. China 
}

\author{S. Nandi}

\affiliation{Department of Physics and Oklahoma Center for High Energy Physics,
Oklahoma State University, Stillwater OK 74078-3072, USA}

\author{Santosh Kumar Rai}

\affiliation{Harish-Chandra Research Institute and Regional Centre for
Accelerator-based Particle Physics, Chhatnag Road, Jhusi, Allahabad 211019, India}

%\date{\today}

\begin{abstract}

The resonant excesses around 2 TeV reported by the ATLAS Collaboration can be explained
in the left-right model, and the tight constraints from
lepton plus missing energy searches can be evaded if the $SU(2)_R$ gauge symmetry is
leptophobic. We for the first time propose an anomaly free leptophobic left-right model
with gauge symmetry $SU(3)_C\times SU(2)_L \times SU(2)_R \times U(1)_{X}$ where
the SM leptons are singlets under $SU(2)_R$.
The gauge anomalies are cancelled by introducing extra vector-like quarks.
The mass of  $Z'$ gauge boson, which cannot be leptophobic, is assumed to be around or above 2.5 TeV
so that the constraint on dilepton final state can be avoided.
Moreover, we find that the $W'\to WZ$ channel cannot explain the ATLAS diboson excess due to
the tension with the
constraint on $W'\to jj$ decay mode. We solve this problem by considering the
mixings between the SM quarks and vector-like quarks.
We show explicitly that the ATLAS diboson excess can be explained in the viable parameter space
of our model, which is consistent with all the current experimental constraints.

\end{abstract}


\preprint{ HRI-RECAPP-2015-019; OSU-HEP-15-07}

%\preprint{ACT-06-09, MIFP-09-23}

\maketitle

\section{Introduction}

Both the ATLAS and CMS Collaborations have performed searches for the massive resonances decaying 
into a pair of weak gauge bosons via the jet substructure techniques, {\it i.e.},
the $pp \to V_1 V_2 \to 4j$ ($V_{1,2}=W^\pm$ or $Z$) channels~\cite{Aad:2015owa, Khachatryan:2014hpa, Khachatryan:2014gha}.
With 20.3 ${\rm fb}^{-1}$ of data at 8 TeV LHC beam collision energies,
the  ATLAS Collaboration have found excesses for narrow width resonances around $2$~TeV in the $WZ$, $WW$, and $ZZ$ channels
with local signal significances of 3.4$\sigma$, 2.6$\sigma$, and
2.9$\sigma$, respectively~\cite{Aad:2015owa}. Moreover, the CMS Collaboration have done the similar searches,
though did not distinguish between $W$- and $Z$-tagged jets, uncovering a $1.4\sigma$ excess near 1.9 TeV~\cite{Khachatryan:2014hpa}.
Interestingly, the CMS Collaboration also reported about 2$\sigma$ and 2.2$\sigma$ excesses near 1.8 TeV
and 1.8--1.9 TeV in the dijet resonance channel and the $e\nu b {\bar b}$ channel, respectively,
which could be explained by a $W' \to W h$ process~\cite{Khachatryan:2015bma, CMS-Preprint}.
Although they are not yet statistically significant, these anomalous events were interpretated
as new physics beyond the Standard Model (SM) due to the correlations among different searches.
Recently, these diboson excesses have been extensively studied~\cite{Fukano:2015hga, Hisano:2015gna, Cheung:2015nha, Dobrescu:2015qna,Alves:2015mua, Gao:2015irw, Thamm:2015csa, Brehmer:2015cia, Cao:2015lia, Cacciapaglia:2015eea, Abe:2015jra,Allanach:2015hba, Abe:2015uaa, Carmona:2015xaa, Chiang:2015lqa, Cacciapaglia:2015nga, Fukano:2015uga, Sanz:2015zha,Chen:2015xql, Omura:2015nwa, Anchordoqui:2015uea, Chao:2015eea, Bian:2015ota, Kim:2015vba, Lane:2015fza, Faraggi:2015iaa,Low:2015uha, Liew:2015osa, Terazawa:2015bsa, Arnan:2015csa, Niehoff:2015iaa, Fichet:2015yia,Petersson:2015rza, Deppisch:2015cua, Aguilar-Saavedra:2015rna,Bian:2015hda,Dev:2015pga,Franzosi:2015zra,
Li:2015yya, Fritzsch:2015aca, Dobado:2015hha, Zheng:2015dua, Llanes-Estrada:2015hfa, Chen:2015cfa, Aydemir:2015nfa, Bandyopadhyay:2015fka, Dobado:2015nqa, Arbuzov:2015pea, Sierra:2015zma, Dobado:2015goa, Ko:2015uma, Collins:2015wua, Allanach:2015blv, Dobrescu:2015jvn, Bhattacherjee:2015svr, Sajjad:2015urz, Wang:2015sxe,
Alves:2015vob, Appelquist:2015vdl, Allanach:2015gkd, Cao:2015zop, Feng:2015rzn}.

The resonances, which are around 2~TeV and have widths less than about 100~GeV, can address
the  ATLAS diboson excess. Because such narrow resonances might suggest new weakly interacting particles,
we will consider the perturbative theories here. For the ATLAS excesses
in the $WZ$, $WW$, and $ZZ$ channels, the tagging selections for each mode used in the analyses are
rather incomplete, and these channels share about 20\% of the events. Thus, it may be difficult to claim
that a single resonance is responsible for all excesses, although such possibility exists.
The reference ranges of
the production cross-section times the decay branching ratio for the 2 TeV resonances
in the $WZ$, $WW$, and $ZZ$ channels are approximately  $4-8$~fb, $3-7$~fb, and $3-9$~fb, respectively.
So, we shall consider that the prefered production cross-section times the decay branching is
from 5 to 10 fb. 

We shall employ the left-right models to explain the ATLAS diboson excess, which have been studied
recently by quite a few groups as well
(For example, see Refs.~\cite{ Dobrescu:2015qna, Gao:2015irw, Cao:2015lia, Dev:2015pga}).
To evade the tight constraints from
lepton plus missing energy searches, we shall consider the leptophobic $SU(2)_R$ gauge symmetry. However,
in the previous studies of such kind of models, anomaly cancellations have not been considered.
In this paper, we for the first time propose an anomaly free leptophobic left-right model
with gauge symmetry $SU(3)_C\times SU(2)_L \times SU(2)_R \times U(1)_{X}$ where
the SM leptons are singlets under $SU(2)_R$.
To cancel the gauge anomalies, we introduce additional vector-like quarks. Because
the $Z'$ gauge boson cannot be leptophobic, we assume its mass to be around or above 2.5 TeV
so that the constraint on dilepton final state can be escaped.
Moreover, we find that the $W'\to WZ$ channel cannot explain the ATLAS diboson excess due to
the tension with the
constraint on $W'\to jj$ decay mode. Interestingly, this problem can be solved via the
mixings between the SM quarks and vector-like quarks.
We show explicitly that the ATLAS diboson excess can be generated in the viable parameter space
of our model, which is consistent with all the current experimental constraints.

This paper is organized as follows. In Section II, we present the anomaly free leptophobic left-right
symmetric model. In Section III, we study the ATLAS diboson excess and other phenomenological
constraints in details.
Our conclusion and summary are given in Section IV.

\section{The Anomaly Free Leptophobic Left-Right Symmetric Model}

It was observed that in a leptophobic left-right symmetric model, one can indeed explain the 
diboson excesses within $2\sigma$ confidence level~\cite{Gao:2015irw, Cao:2015lia}. However,
in order to escape the constraint from the $Z'$ decay $Z' \to \ell^+ \ell^-$, one needs to consider 
the leptophobic $SU(2)_R$ model. Earlier models along this line have not been free of anomalies. 
In order to cancel the gauge anomalies,  we consider a similar model by introducing 
extra vector-like quarks in the theory. 
%We shall propose two kinds of the models: Model I with vector-like quarks, while Model II with vector-like leptons.

In the left-right symmetric model, the gauge symmetry is 
$SU(3)_C\times SU(2)_L \times SU(2)_R \times U(1)_{X}$ where $X\equiv B-L$
in the original model~\cite{Pati:1975, Mohapatra:1974, Senjanovic:1975, Senjanovic:1978, Mohapatra:1979, Mohapatra:1980} . 
We consider that the $ SU(2)_R \times U(1)_{X}$ gauge symmetry
is broken down to  $U(1)_Y$ by a doublet Higgs field $H'$ around the TeV scale, and
the $SU(2)_L\times U(1)_Y$ gauge symmetry is further broken down
to $U(1)_{EM}$ via a bidoublet Higgs field $\Phi$ and a Higgs doublet $H$. To cancel the gauge
anomalies, we introduce the extra quarks $XQ^{Rc}_i$, $XU^c_i$, and $XD^c_i$,
%and the extra leptons $XL^{Rc}_i$, $XE_i$ and $XN_i$ in Model II, 
which become vector-like particles after the $SU(2)_R\times U(1)_X$ gauge symmetry breaking.
As in the supersymmetric SMs, we denote the SM left-handed quark doublets,
right-handed up-type quarks, right-handed down-type quarks, left-handed
lepton doublets and right-handed charged leptons  as  
%and neutrinos as
$Q_i$, $U_i^c$, $D_i^c$, $L_i$ and $E_i^c$, 
%and $N_i^c$, 
respectively. 
We also define $Q^R_i\equiv (U_i^c, ~D_i^c)$ and $XQ^{Rc}_i\equiv (XU_i, ~XD_i)$. 
Thus,  $(XU_i, ~XU_i^c)$/$(XD_i, ~XD_i^c)$  will form the vector-like quarks after the symmetry breaking. 
We present the particles and their quantum numbers in Table~\ref{Particle-Spectrum-I}. 
%\ref{Particle-Spectrum-II}, respectively. 
It is easy to show that the model is anomaly free. By the way, instead of $H'$,
we can also introduce an $SU(2)_R$ triplet Higgs field with $U(1)_X$ charge one to break 
the $SU(2)_R\times U(1)_X$ gauge symmetry down to $U(1)_{EM}$.
However, such triplet Higgs field cannot give the vector-like masses to vector-like particles after gauge 
symmetry breaking. So we will not consider this kind of scenarios here.

\begin{table}[h]
\begin{tabular}{|c|c|c|c|c|c|}
\hline
~$Q_i$~ & ~$(\mathbf{3}, \mathbf{2}, \mathbf{1}, \mathbf{1/6})$~ &
$Q^R_i$ &  ~$(\mathbf{\overline{3}}, \mathbf{1}, \mathbf{2}, \mathbf{-1/6})$~ &
~$L_i,~H$~ & ~$(\mathbf{1}, \mathbf{2}, \mathbf{1}, \mathbf{-1/2})$~\\
\hline
$E_i^c$ &  $(\mathbf{1}, \mathbf{1},  \mathbf{1}, \mathbf{1})$ &
$XQ_i^{Rc}$ &  ~$(\mathbf{3}, \mathbf{1},  \mathbf{2}, \mathbf{1/6})$~ &
~$XU_i^c$~ &  ~$(\mathbf{\overline{3}}, \mathbf{1},  \mathbf{1}, \mathbf{-2/3})$~ \\
\hline
~$XD_i^c$~ &  $(\mathbf{\overline{3}}, \mathbf{1},  \mathbf{1}, \mathbf{1/3})$ &
$\Phi $ &  ~$(\mathbf{1}, \mathbf{2}, \mathbf{2}, \mathbf{0})$~&
$H'$ &  ~$(\mathbf{1}, \mathbf{1}, \mathbf{2}, \mathbf{-1/2})$~   \\
\hline
\end{tabular}
\caption{The particles and their quantum numbers under the
  $SU(3)_C \times SU(2)_L \times SU(2)_R \times U(1)_X$ gauge symmetry in Model I.}
\label{Particle-Spectrum-I}
\end{table}

%\begin{table}[h]
%\begin{tabular}{|c|c|c|c|c|c|}
%\hline
%~$Q_i$~ & ~$(\mathbf{3}, \mathbf{2}, \mathbf{1}, \mathbf{1/6})$~ &
%$Q^R_i$ &  ~$(\mathbf{\overline{3}}, \mathbf{1}, \mathbf{2}, \mathbf{-1/6})$~ &
%~$L_i,~H$~ & ~$(\mathbf{1}, \mathbf{2}, \mathbf{1}, \mathbf{-1/2})$~\\
%\hline
%$E_i^c$ &  $(\mathbf{1}, \mathbf{1},  \mathbf{1}, \mathbf{1})$ &
%$XL_i^{R}$ &  ~$(\mathbf{1}, \mathbf{1},  \mathbf{2}, \mathbf{1/2})$~ &
%~$XE_i$~ &  ~$(\mathbf{\overline{1}}, \mathbf{1},  \mathbf{1}, \mathbf{-1})$~ \\
%\hline
%~$XN_i$~ &  $(\mathbf{\overline{1}}, \mathbf{1},  \mathbf{1}, \mathbf{0})$ &
%$\Phi $ &  ~$(\mathbf{1}, \mathbf{2}, \mathbf{2}, \mathbf{0})$~&
%$H'$ &  ~$(\mathbf{1}, \mathbf{1}, \mathbf{2}, \mathbf{-1/2})$~   \\
%\hline
%\end{tabular}
%\caption{The particles and their quantum numbers under the
%  $SU(3)_C \times SU(2)_L \times SU(2)_R \times U(1)_X$ gauge symmetry in Model II.}
%\label{Particle-Spectrum-II}
%\end{table}

The $SU(2)_R \times U(1)_X$ gauge symmetry is broken down to $U(1)_Y$ via the
vacuum expectation value (VEV) of $H'$ as below
\beq
H' = \begin{pmatrix} \chi^{\prime 0} \\ \chi^{\prime -}\end{pmatrix}~,~~
\VEV{H'} =
\mfrac{1}{\sqrt{2}}
\begin{pmatrix} u \\ 0 \end{pmatrix}~.
\eeq 
Subsequently the $SU(2)_L\times U(1)_Y$ gauge symmetry is broken down to $U(1)_{EM}$ via
the VEVs of $\Phi$ and $H$ as follows
\begin{align}
  &H = \begin{pmatrix} \chi^0 \\ \chi^-\end{pmatrix},
&\VEV{H} &=
\mfrac{1}{\sqrt{2}}
\begin{pmatrix} v_3 \\ 0 \end{pmatrix}, \nn \\
&\Phi=\begin{pmatrix}\phi_1^{0} & \phi_1^{+} \\ \phi_2^{-} & \phi_2^{0} \end{pmatrix},\qquad
& \VEV{\Phi} & = \mfrac{1}{\sqrt{2}}
\begin{pmatrix} v_1 & 0 \\ 0 & v_2 \end{pmatrix}.
\end{align}
As we shall discuss below, $H$ will give masses to the charged leptons.  For simplicity,
we assume $v_1 >> v_3$ and $v_2 >> v_3$ such that $v = \sqrt{v_1^2 + v_2^2+v_3^2} 
 \simeq \sqrt{v_1^2 + v_2^2}$.  Note that as $W'$ and $Z'$ are around 2-3 TeV, we have 
 $v << u$. So we define a small parameter, $1/x$ where
\be
x \,\, \equiv \,\, \frac{u^2}{v^2} \,\,,
\ee
with $x \, \gg \, 1$ and a mixing angle $\beta \equiv \arctan(v_1/v_2)$.

The SM fermion Yukawa terms in both models are
\begin{eqnarray}
  -{\cal L} &=& y_{ij}^Q Q_i Q_j^R \Phi ~~ + ~~  y_{ij}^L  L_i E^c_j H ~,~\,
  \label{eq:smyukawa}
\end{eqnarray}
where $y_{ij}^Q$ and $y_{ij}^L$ are Yukawa couplings, and $i/j=1,~2,~3$. Thus, as in the left-right
symmetric models with only one bi-doublet, the SM quark masses and CKM mixings
are still a problem, which will be discussed elsewhere.

The additional Yukawa and bilinear mass terms in our model are
\begin{eqnarray}
  -{\cal L} &=& y_{ij}^{QXU} Q_i XU_j^c {\tilde H} ~~ + ~~ y_{ij}^{QXD}  Q_i XD_j^c H
%  + y^{UXQ}_{ij} XQ_i^{Rc} U_j^c {\tilde H'} 
%+ y^{DXQ}_{ij} XQ_i^{Rc} D_j^c H'
  ~~ + ~~ y^{XQU}_{ij} XQ_i^{Rc} XU_j^c {\tilde H'}
  \nonumber \\ &&
  + ~~ y^{XQD}_{ij} XQ_i^{Rc} XD_j^c H' ~~ + ~~ \mu_{ij} Q_i^R XQ_j^{Rc} 
  + h.c. ~,~ 
    \label{eq:bsmyukawa}
\end{eqnarray}
where ${\tilde H} \equiv i \sigma_2 H^*$, ${\tilde H'} \equiv i \sigma_2 H^{\prime *}$,
$y_{ij}^{QXU}$, $y_{ij}^{QXD}$, 
%$y^{UXQ}_{ij}$, $y^{DXQ}_{ij} $,
$y^{XQU}_{ij}$ and $y^{XQD}_{ij} $ 
are Yukawa couplings, and $\mu_{ij}$ are bilinear mass parameters.
The $y^{XQU}_{ij}$ and $y^{XQD}_{ij} $ terms will give the masses
to the vector-like particles $XQ_i^{Rc}$ and $XU^c_i/XD^c_i$.
Interestingly, after we integrate out the vector-like particles,
we have new quark Yukawa terms and they 
may explain the SM quark masses and CKM mixings, which will be studied elsewhere.

%The extra Yukawa and bilinear mass terms in Model II are
%\begin{eqnarray}
%  -{\cal L} &=& y_{ij}^{LXL} L_i XL_j^R \Phi + y_{ij}^{XLE} XL_i^R XE_j {\tilde H'}
%  + y_{ij}^{XLN} XL_i^R XN_j H' + \mu_{ij} E_i^c XE_i~, ~\,
%\end{eqnarray}
%where $y_{ij}^{LXL}$, $y_{ij}^{XLE}$ and $y_{ij}^{XLN}$ are Yukawa couplings, and
%$\mu_{ij}$ are bilinear mass parameters. The $y^{XLE}_{ij}$ and $y^{XLN}_{ij} $ terms
%will give the masses to the vector-like particles $XL_i^{R}$ and $XE_i/XN_i$.

The gauge couplings $g_L$, $g_R$, and $g_X$ respectively
for $SU(2)_L$, $SU(2)_R$, and $U(1)_X$ are given by
\be
g_L = \frac{e}{\sin\theta_W}, \,\,\,\, g_R = \frac{e}{\cos\theta_W \sin{\phi}}, \,\,\,\,
g_X = \frac{e}{\cos\theta_W \cos{\phi}} \,\,.
\label{eq:gcoup}
\ee
where  $e$ is the $U(1)_{EM}$ gauge coupling,  $\theta_W$ is the weak mixing angle, and
 the mixing angle $\phi \equiv \arctan(g_X/g_R)$.

We denote $SU(2)_L \times SU(2)_R \times U(1)_X$ gauge bosons as follows
\begin{align*}
SU(2)_L: ~W_{1,\mu}^{\pm}, ~W_{1,\mu}^{3}~;~~~
SU(2)_R: ~W_{2,\mu}^{\pm}, ~W_{2,\mu}^{3}~;~~~
U(1)_{X}: ~X_{\mu}~.~
\end{align*}
After gauge symmetry breaking, 
the mass eigenstates of the charged and neutral gauge bosons at the order of $1/x$  are
\begin{eqnarray}
W_\mu^\pm &=& {W_1^\pm}_\mu +\frac{\sin\phi \sin2\beta}{x
  \tan\theta_W}{W_2^\pm}_\mu \, ,
\nn \\
{W^\prime}_\mu^\pm &=&  -\frac{\sin\phi \sin2\beta}{x \tan\theta_W}
{W_1^{\pm}}_{\mu}+{W_2^{\pm}}_{\mu}  \, , \nn \\
{A    }_{\mu} &=& \sin\theta_W {W_1^3}_{\mu} +\cos\theta_W (\sin\phi {W^3_2}_{\mu} + \cos\phi X_{\mu})  \, , 
\nn \\
Z_\mu &=& {W_Z^3}_\mu +\frac{\sin \phi \cos^3 \phi}{x\sin \theta_W}
                           {W_H^3}_\mu\, , \nn \\
Z_\mu^{\prime} &=&  -\frac{\sin\phi\cos^3\phi}{x\sin\theta_W}
{W_Z^3}_\mu + {W_H^3}_\mu \, ,
\label{Eq-W}
\end{eqnarray}
where $W_H^3$ and $W_Z^3$ are 
\begin{eqnarray}
{W_H^3}_{\mu} &=& \cos\phi   {W_2^3}_{\mu} - \sin\phi X_{\mu}\,, \nn \\
{W_Z^3}_{\mu} &=& \cos\theta_W {W_1^3}_{\mu} -\sin\theta_W (\sin\phi {W_2^3}_{\mu} + \cos\phi X_{\mu})\, .
\label{Eq-Z}
\end{eqnarray}
The corresponding masses for $W^{\prime}$ and $Z^{\prime}$ gauge bosons are 
\bea
M_{{W^{\prime}}^{\pm}}^{2} = \frac{e^{2}v^{2}}{4\cos^{2}\theta_W \sin^{2}{\phi}}\left(x+1\right)\,,
\quad
M_{Z^{\prime}}^{2}  = \frac{e^{2}v^{2}}{4\cos^{2}\theta_W \sin^{2}{\phi}\cos^{2}{\phi}}\left(x+\cos^{4}
{\phi}\right)\,.
\label{mzp_bp1}
\eea

The relevant Feynman rules of the gauge-fermion couplings for the $SU(2)_L$ doublets ($P_L$),
$SU(2)_R$ doublets ($P_R$),
and $SU(2)_L\times SU(2)_R$ singlets ($P_S$) are given by 
\begin{equation}
  W^{\prime\pm} \overline{f}f':\quad \frac{e}{\sqrt{2} \sin\theta_W} \left(f_{W'L} P_L + f_{W'R} P_R + f_{W'S} P_S \right)~,
  \label{FM-W}
\end{equation}	
with
\begin{eqnarray}
  f_{W'L} = -\frac{\sin\phi \sin(2\beta)}{x \tan\theta_W}~, \quad \quad f_{W'R} = \frac{\tan\theta_W}{\sin\phi}~,~~
  f_{W'S} = 0 ~,~
\end{eqnarray}
and 
\begin{equation}
Z^{\prime} \overline{f} f: \quad \frac{e}{\sin\theta_W \cos\theta_W} \left(f_{Z'L} P_L + f_{Z'R} P_R + f_{Z'S} P_S\right)~,
\label{eqn:Zprime-ffbar}
\end{equation}	
with
\begin{eqnarray}
f_{Z'L} &=& (T_L^3-Q) \sin\theta_W\tan\phi-(T_L^3-Q\sin^2\theta_W)\frac{\sin\phi \cos^3\phi}{x \sin\theta_W}~,~ \\
f_{Z'R} &=& (T_R^3-Q\sin^2\phi) \frac{\sin\theta_W}{\sin\phi\cos\phi} +Q\frac{\sin\theta_W\sin\phi\cos^3\phi}{x}~,~ \\
f_{Z'S} &=& -Q \sin\theta_W\tan\phi + Q\frac{\sin\theta_W\sin\phi\cos^3\phi}{x}~.~ 
\label{eqn:Zprime-LR}
\end{eqnarray}

With all out-going momenta, the gauge boson self-couplings are given as follows.
The three-point couplings are
\begin{equation}
	V_1^{\mu}(k_1) V_2^{\nu}(k_2) V_3^{\rho}(k_3): \ \ 
	- i f_{V_1V_2V_3} \left[ g^{\mu\nu}(k_1-k_2)^{\rho}
	+ g^{\nu\rho}(k_2-k_3)^{\mu}
	+ g^{\rho\mu}(k_3-k_1)^{\nu} \right]~,
\end{equation}
where the coupling strengths $f_{V_1V_2V_3}$ for the $WWZ'$ and $W'WZ$ are
\begin{eqnarray}
 f_{WWZ'} = \frac{e \sin\phi \cos^3\phi \cot\theta_W}{x \sin\theta_W}~,\quad
 f_{W'WZ} = \frac{e \sin\phi \sin(2\beta)}{x \sin^2\theta_W}~.
 \label{Zprime-ww}
\end{eqnarray}

%\textcolor{red}{
We note that the number of physical scalar fields in the model after symmetry breaking is quite large 
due to the presence of several scalar multiplets \footnote{The details of the full scalar spectrum and
its implications will be discussed elsewhere as it would not play a significant role in the explanation
of the diboson excess.}. We work in the approximation that of the three remaining
CP-even neutral Higgs fields, we assume that we can decouple one of them by appropriate choice 
of the bare parameters in the scalar potential. Therefore, we assume that only two low-lying 
CP-even states will have significant mixing. Expanding the Higgs field $\Phi$ around the vacuum we have, 
\begin{align*}
\Phi=\begin{pmatrix} \frac{v_1+h_1}{\sqrt{2}} & \phi_1^{+}\\ \phi_2^{-} & \frac{v_2+h_2}{\sqrt{2}}\end{pmatrix}.
\end{align*}
The CP even states would mix and we can write them in terms of the physical basis as 
\begin{equation}
 \begin{pmatrix} h_1 \\ h_2 \end{pmatrix} = \begin{pmatrix} \cos\alpha & \sin\alpha \\ -\sin\alpha & \cos\alpha \end{pmatrix}
 \begin{pmatrix} h \\ h^{\prime} \end{pmatrix}                     
\end{equation}
In the decoupling limit where $\alpha = \beta - \frac{\pi}{2}$ we can rewrite the states in terms 
of the angle defined by the ratio of the two doublet vevs only :   
\begin{equation}
 \begin{pmatrix} h_1 \\ h_2  \end{pmatrix} =  \begin{pmatrix} h\sin\beta - h' \cos\beta \\ h\cos\beta + h^{\prime}\sin\beta \end{pmatrix}
\end{equation}
We identify $h$ as the SM Higgs 
%	} \\
such that the $hWW'$ and $hZZ'$ couplings are given by 
\begin{eqnarray}
 hWW': g^{\mu\nu}\frac{e^2 v}{2\sin^2\theta_W}  f_{HWW'}~, \quad
  hZZ':  g^{\mu\nu}\frac{e^2 v}{2\sin^2\theta_W\cos^2\theta_W}  f_{HZZ'}~,
\end{eqnarray}
where the coupling strengths are
\begin{eqnarray}
  f_{hWW'} &=& -\frac{\sin(2\beta)\tan\theta_W}{\sin\phi}
  +\frac{\sin(2\beta)(\tan\theta_W-\cot\theta_W\sin^2\phi)}{x \sin\phi}~,\\
  f_{hZZ'} &=& -\frac{\sin\theta_W}{\tan\phi}
  +\frac{\cos^3\phi( \sin^2\theta_W \cos^2\phi -\sin^2\phi )}{x \sin\theta_W \sin\phi}~.
  \label{FM-Z}
\end{eqnarray}
We now have all the relevant interaction strengths in the model to study the phenomenology 
of the additional heavy gauge bosons in the theory. In the next section we consider their production at the
LHC and how the diboson excess can be explained in our model.

%%%%%%%%%
\section{The Consistent Diboson Excess Analyses}
\begin{figure*}[h]
\centering
\begin{minipage}[b]{.45\textwidth}
\includegraphics[height=0.8\linewidth,width=1.0\linewidth]{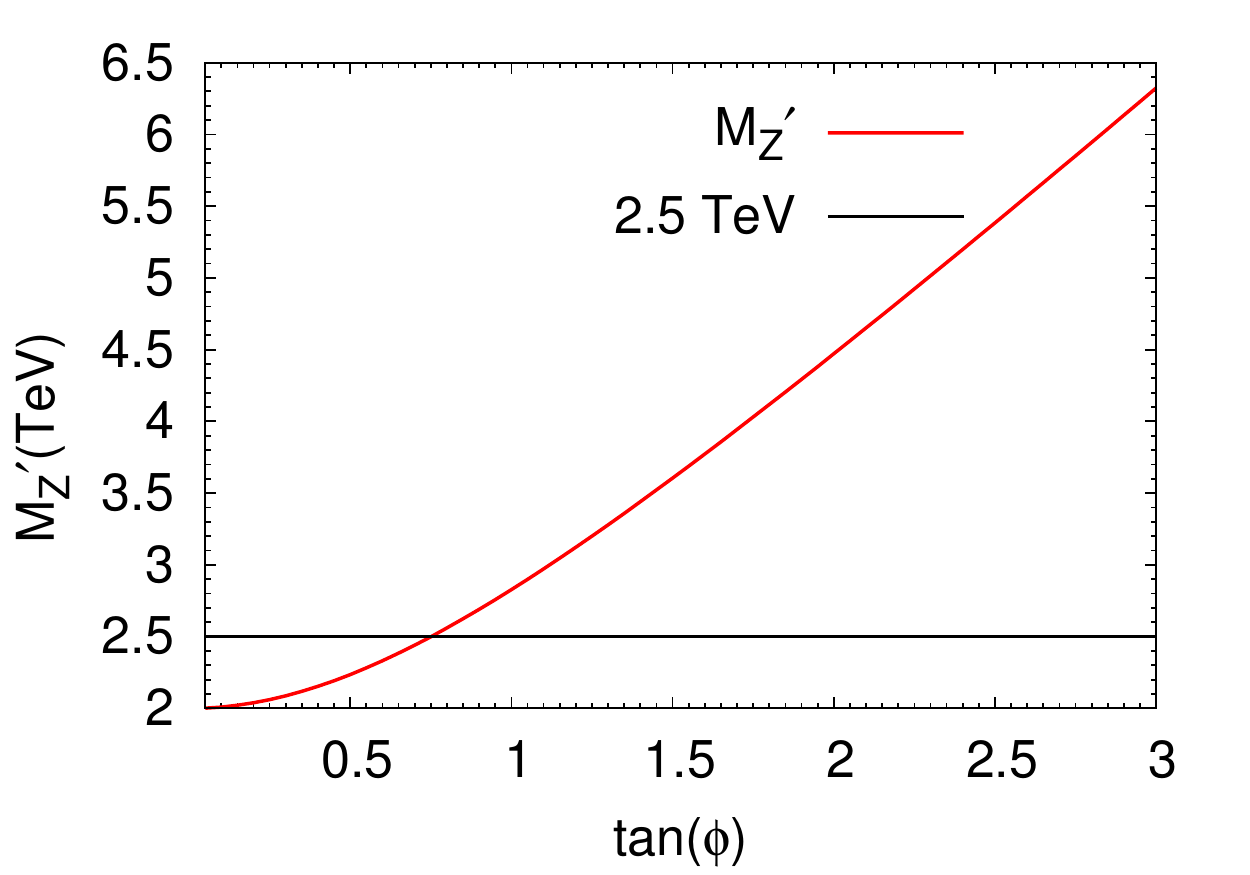}
\caption{Mass of $Z'$ versus $\tan\phi$.}\label{fig:Zpmass}
\end{minipage}\qquad
\begin{minipage}[b]{.45\textwidth}
\includegraphics[height=0.8\linewidth,width=1.0\linewidth]{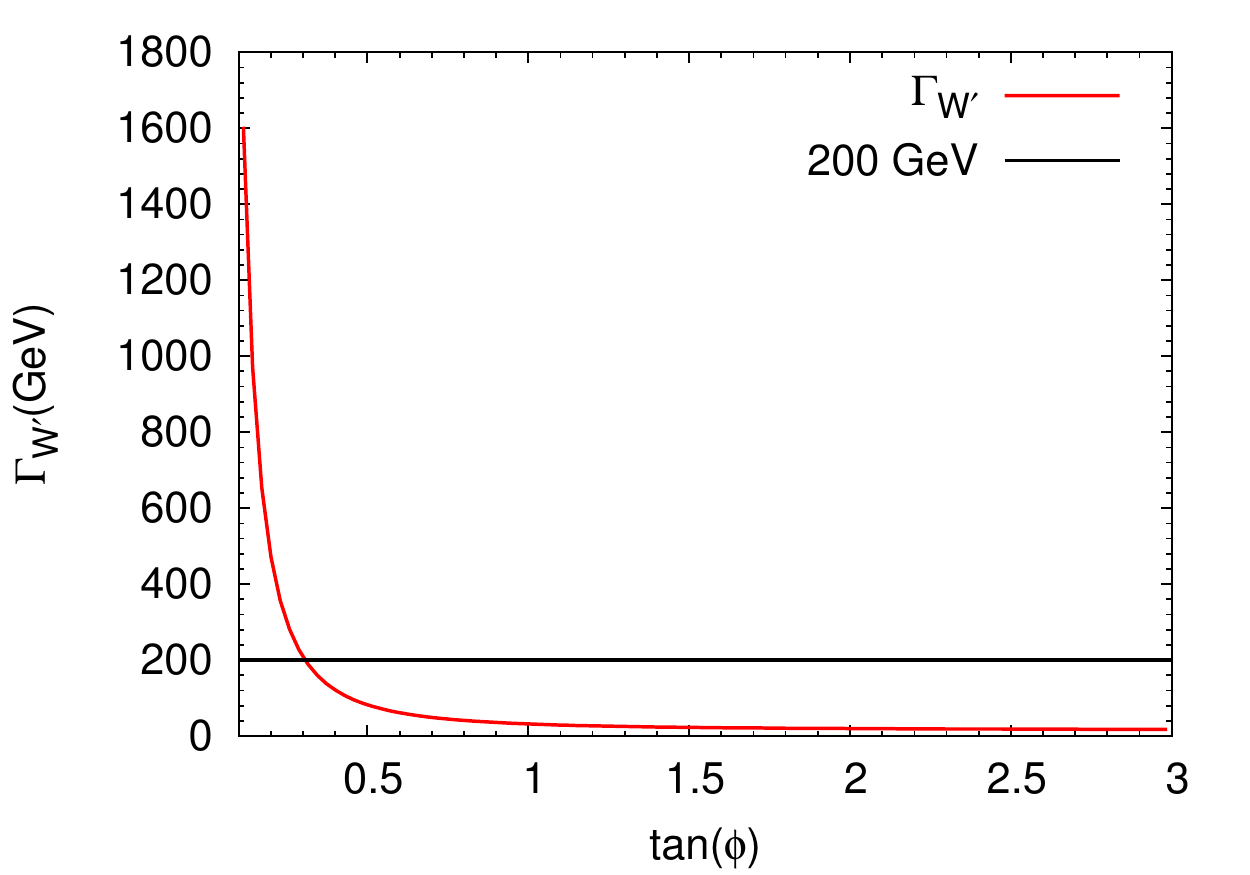}
\caption{Width of W$^{\prime}$ versus $\tan\phi$.}\label{fig:Wpwidth}
\end{minipage}
\end{figure*}

%%%%%%%%%%
The model we propose gives us the heavy gauge bosons $W'^\pm$ and $Z'$ with their
masses depending on the $H'$ doublet Higgs VEV ($u$) and $SU(2)_R$ gauge coupling.
We note that the diboson excess reported by ATLAS could be the combined contributions from 
the $W'$ and $Z'$ productions decaying to a pair of electroweak gauge bosons. However,  
a strong constraint on the dilepton final state suggests that the $Z'$ production with mass of 
around 2 TeV could be in contradiction to it unless the $Z'$ is leptophobic as well. 
So we first try and set the parameter space such that the $Z'$ is too heavy and evades such 
bounds. We shall come back to the $Z'$ phenomenology to see if such a requirement is at all 
necessary in our model. For simplicity we shall set the mass of $Z'$ to be 2.5 TeV and 
above.  The relevant parameters that need to be considered in the analysis would be the 
gauge couplings, $g_L, g_R$ and $g_X$ which can be re-parameterised in terms of the known 
parameters, {\it viz.} electric charge $e$, Weinberg angle $\theta_W$, and an 
additional unknown angle $\phi$ as shown in Eq.~(\ref{eq:gcoup}). As we would like to 
achieve a signal for a 2 TeV resonance, the $M_{W'}$ is fixed at that value which in 
turn makes the ratio $x=u^2/v^2$ dependent on $\phi$. Thus, as we change $\phi$ the 
VEV $u$ also changes and affects the $Z'$ mass. We plot the dependence of the $Z'$ mass on the 
value for $\tan\phi$ for a fixed $M_{W'}=2$ TeV in Fig.~\ref{fig:Zpmass}.
As the $Z'$ mass is found to increase with increasing values of $\tan\phi$,  the 
mass of Z$^\prime$ becomes greater than 2.5 TeV for $\tan\phi$ greater than 0.76. 

We also want the 2 TeV resonance to have a narrow width which we achieve by demanding 
the total decay width $\Gamma$ of the resonance to satisfy $\Gamma/M_V \leq 0.1$. Since 
the resonant signal must now come only from the $W'$, we plot  the dependence of the 
width of W$^{\prime}$ on  $\tan\phi$ in Fig.~\ref{fig:Wpwidth} where again 
the mass of $W'$ has been set to 2 TeV. It is clear that narrow width approximation is 
valid for a 2 TeV $W'$ only when $\tan\phi \geq 0.31$. 
%%%%%%%
%%%%%

We now turn our attention to the signal and the parameters that affect the rates, 
since the on-shell production of the $W'$ depends on the values of $\phi$ and $\beta$. Note that the
interaction strengths of the $W'$ to the SM particles suggests that the
decay of $W' \to WZ$ can be maximized for $\tan\beta=1$ while the production of the right-handed gauge 
boson is enhanced for smaller values of $\tan\phi$. Therefore, throughout the analysis we have kept 
$\tan{\beta}=1$ and varied $\tan\phi$ as it governs the production of $W'$ as well as its dominant decay channels.
We take the range of $\tan\phi$ from 0.75 to 3 to find a viable parameter space in the model 
which can produce the  signal rates for the diboson excess as reported by the ATLAS Collaboration, while satisfying 
constraints in other channels. 

%%%%%%%%
\begin{figure*}[h]
\centering
\begin{minipage}[b]{.45\textwidth}
\includegraphics[height=0.8\linewidth,width=1.0\linewidth]{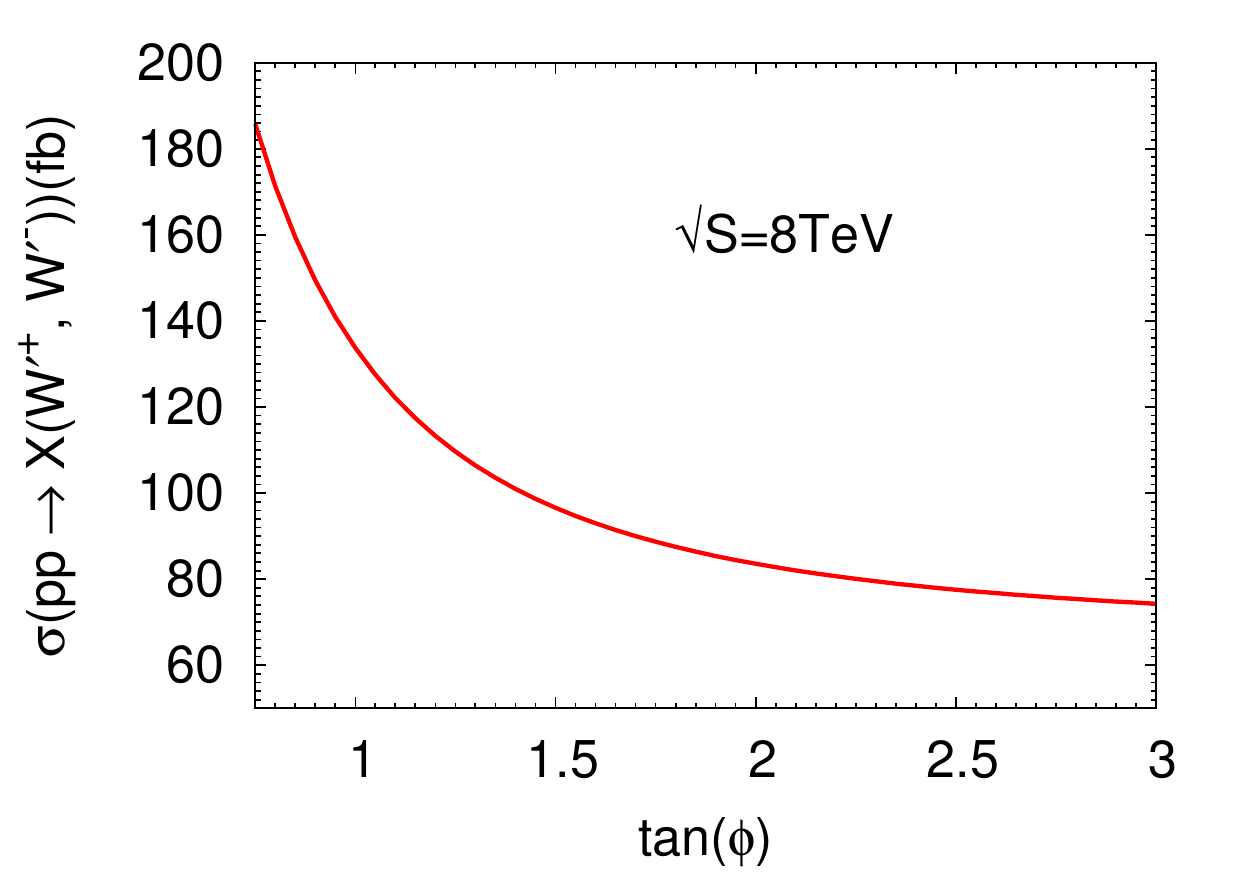}
\caption{Production cross section of $W'$ versus $\tan\phi$.}\label{fig:Wpcross}
\end{minipage}\qquad
\begin{minipage}[b]{.45\textwidth}
\includegraphics[height=0.8\linewidth,width=1.0\linewidth]{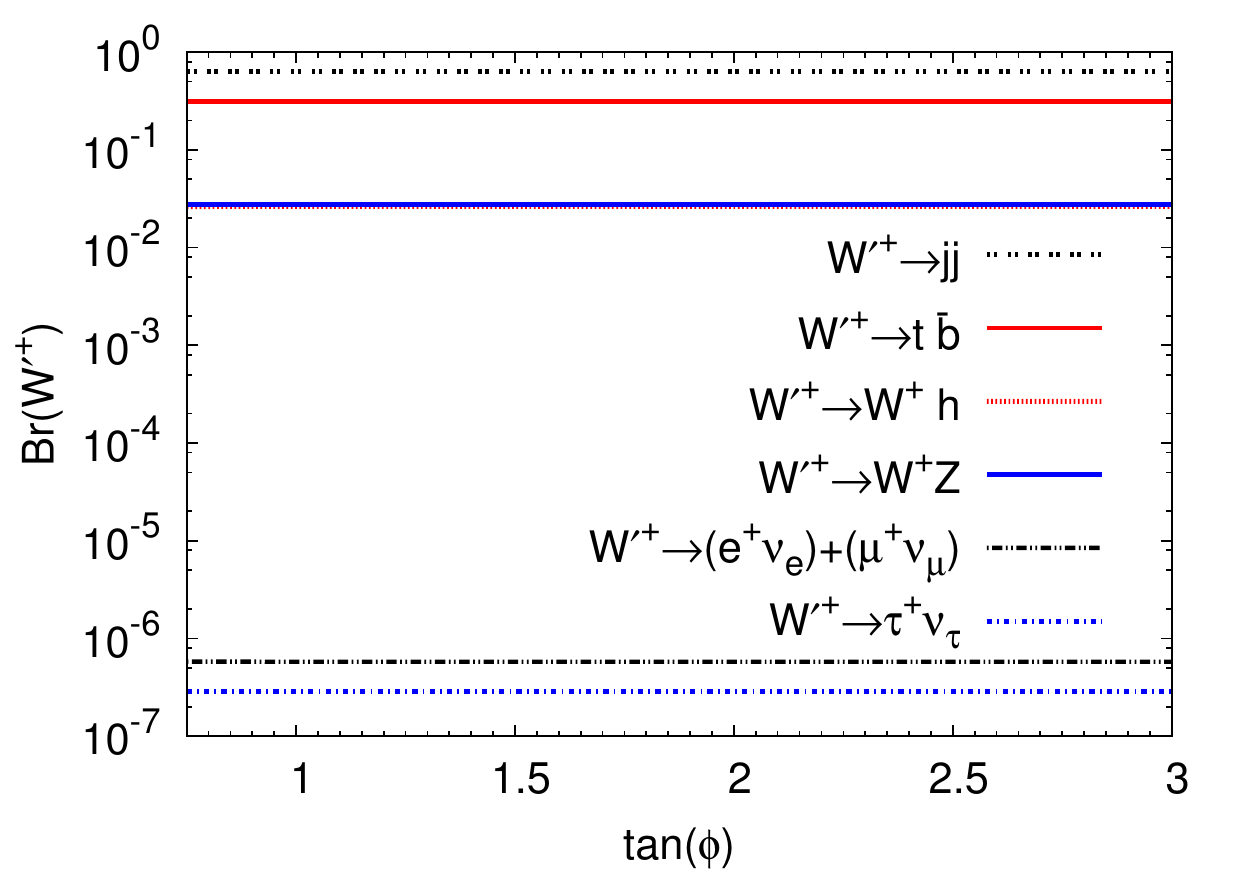}
\caption{Branching ratio for different decay channels for  W$^{\prime}$ versus $\tan\phi$.}
\label{fig:BrWp}
\end{minipage}
\end{figure*}
%%%%%

In Fig.~\ref{fig:Wpcross} we plot the production cross section of the $W'$ as a function of $\tan\phi$. 
The interaction strength  of the $W'$ to the quarks become weaker as $\tan\phi$ increases which therefore 
leads to a drop in the production cross section of the $W'$. 
Finally, the final channels observed would depend on how the $W'$ decays and we show this in 
Fig.~\ref{fig:BrWp}.  As the model by construction makes the $W'$ leptophobic, the branching ratios for 
$W'$ decaying to leptons and neutrinos are negligible. However, the right-handed gauge boson 
does have a substantial coupling strength to the SM quarks and the new exotic heavy quarks 
which are doublets under the new $SU(2)_R$. 
This not only helps in producing the $W'$ with large cross sections
at the LHC, it also leads to strong signal rates to final states such 
as $jj$ (sum over first two generations of quarks) and $t\bar{b}$ which 
are  constrained by the LHC data. 
We choose the heavy exotic vector-like quarks to have mass
$M_{XQ} \geq M_{W'}/2$ such that the $W'$ decay to a pair of these
exotic quarks is kinematically forbidden.
The constraint on the  decay mode $W'\rightarrow t\overline{b}$ 
is given by~\cite{Aad:2014xea}
\begin{align*}
{\sigma(pp \rightarrow W')\times Br(W'\rightarrow t\overline{b})} \lesssim 120~ fb,
\end{align*}
while the dijet limits for the $jj$ final state is~\cite{Khachatryan:2015sja}
\begin{align*}
{\sigma(pp \rightarrow W')\times Br(W'\rightarrow jj)} \lesssim 100 ~ fb. 
\end{align*}
Another relevant bound is for the $W'\rightarrow Wh$ decay mode~\cite{Khachatryan:2015bma}
\begin{align*}
{\sigma(pp \rightarrow W')\times Br(W'\rightarrow Wh)} \lesssim 7 ~fb. 
\end{align*}
From the equivalence theorem for a spontaneously broken gauge symmetry
we expect the $Wh$ rates to be around the $WZ$ rate and therefore should satisfy the existing 
constraints as the signal is expected to be in the same range.  

\begin{figure}[ht]
 \includegraphics[width=.6\linewidth]{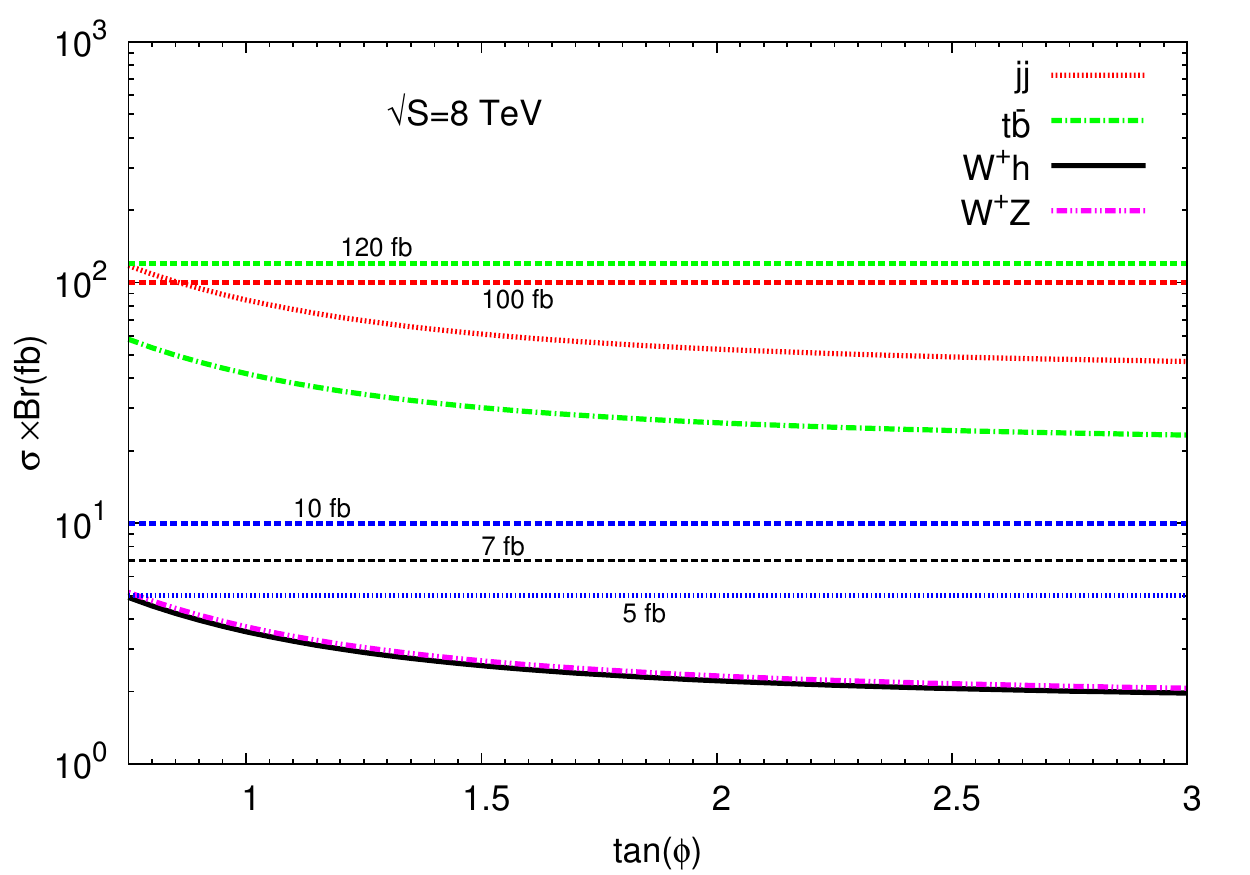}
 \caption{Cross section times branching ratio for  W$^{\prime}$ versus $\tan\phi$.}
 \label{fig:WpXStimesBr}
\end{figure}

We now analyse and see whether the model can satisfy the constraints and give the 
required event rates for the diboson excess. To check this we plot the cross section times 
the branching fractions ($\sigma\times BR$) to different final states in  Fig.~\ref{fig:WpXStimesBr} with 
the corresponding constraints mentioned in tandem. The horizontal lines with fixed numeric values
in the figure represent the respective upper bounds on the final state at the parton level, as 
presented by the LHC run-I experiment.
In the figure we also have illustrated a band with a range of 5-10 fb for the diboson signal which 
could account for and explain the diboson excess, such that the $\sigma\times BR$ for the 
$W' \rightarrow WZ$ should lie within that band. The dijet and top production rates are 
found to be still quite large but they do satisfy the existing constraints for nearly the complete 
range of $\tan\phi$. However, the signal rates for the $WZ$ channel is not adequate to explain 
the diboson excess. The constraint on $W' \rightarrow jj$ decay mode is not 
satisfied for $\tan\phi$ values below $\sim 0.85$. However, up till now, we have 
ignored the fact that there could be significant mixing between the heavy vector-like quarks 
with the right-handed SM quarks. Such mixings are not very strongly constrained by flavor 
physics like the left-handed ones. Once allowed, this would not only lead to a suppression in the 
$W'$ production but will also suppress the decay modes of $W'$ to the quark final states. Such 
a suppression can lead to an enhancement of the $W' \to WZ$ branching and therefore increase the 
rates enough to accommodate the diboson excess. 

%%%%%
\begin{figure*}[h]
\centering
\begin{minipage}[b]{.45\textwidth}
\includegraphics[height=0.9\linewidth,width=1.1\linewidth]{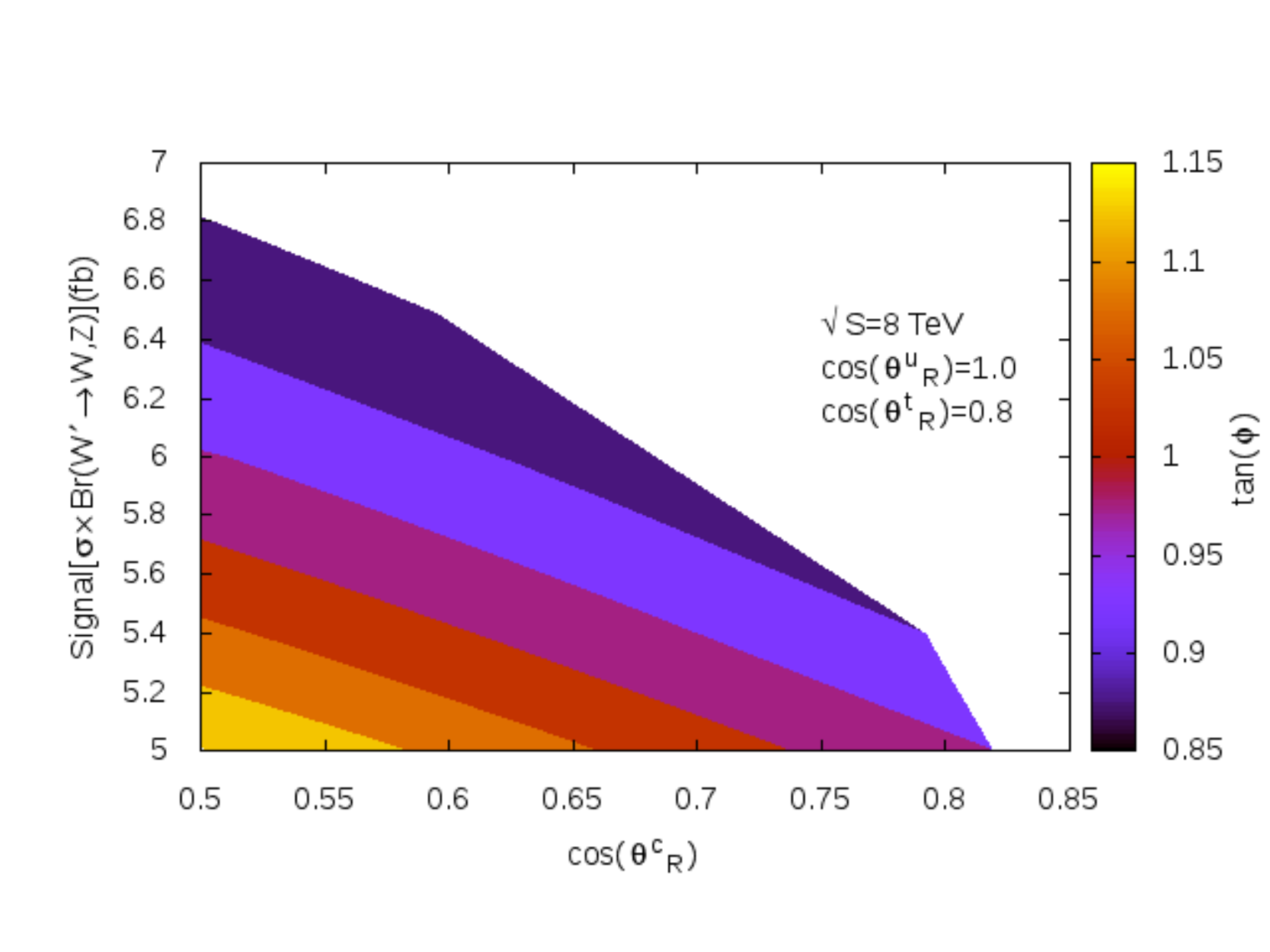}
\caption{Signal for cross section times branching ratio for  $W'\rightarrow WZ$ 
versus  $\cos(\theta_{R}^{c})$ for different values of $\tan\phi$.}\label{fig:ccRswz-8TeV}
\end{minipage}\qquad
\begin{minipage}[b]{.45\textwidth}
\includegraphics[height=0.9\linewidth,width=1.1\linewidth]{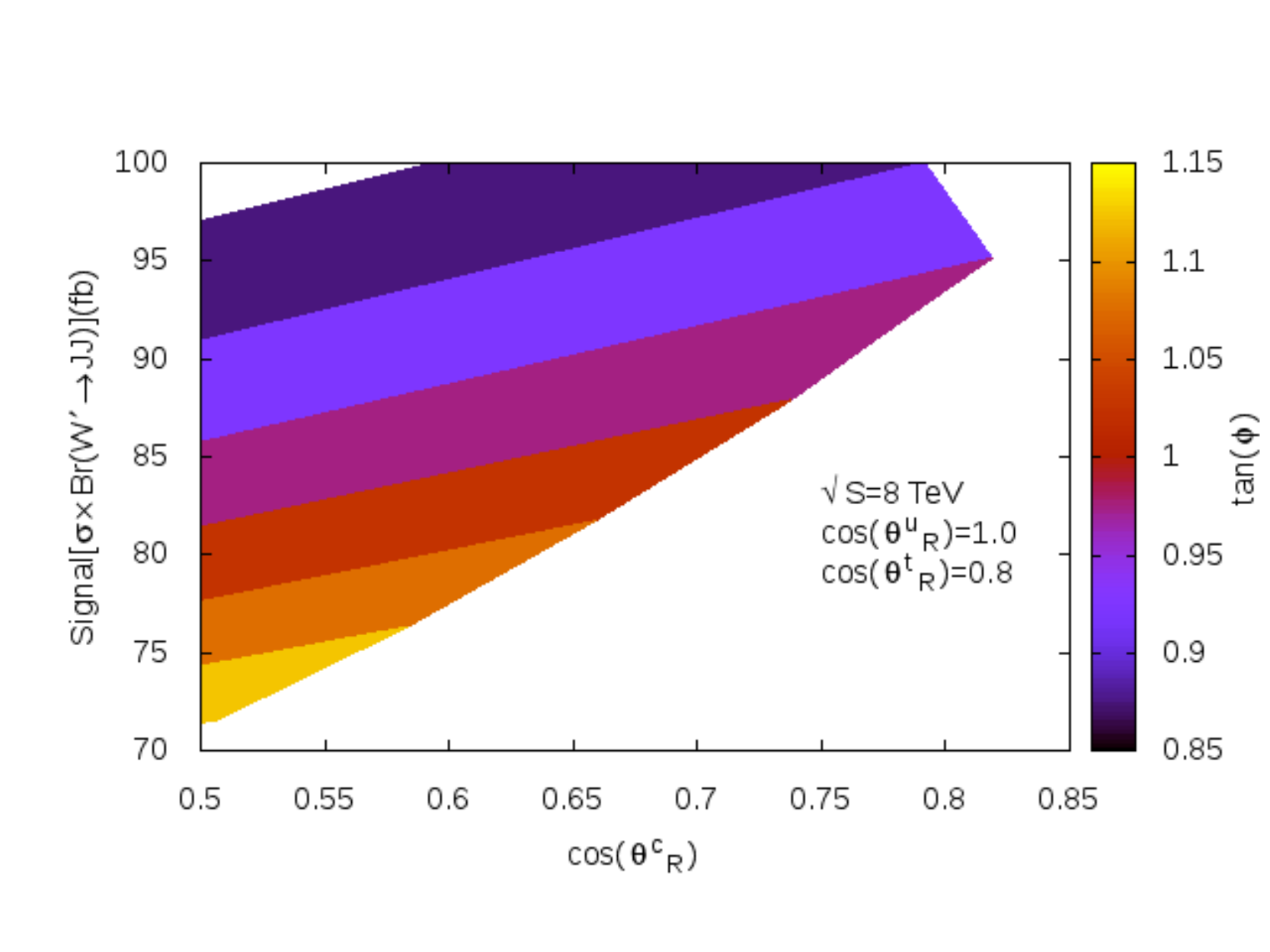}
 \caption{Signal for cross section times branching ratio for  $W^{\prime}\rightarrow jj$ versus $\cos(\theta_{R}^{c})$ for different values of $\tan\phi$.} 
 \label{fig:ccRsJJ-8TeV}
\end{minipage}
\end{figure*}
%%%%%
To achieve {${\sigma(pp \rightarrow W') \times Br(W' \rightarrow WZ)} \sim 5-10$ fb,  we 
include the following mixing between the right-handed quark and the right-handed new heavy 
quark sector~\footnote{We must point out that the masses of the vector-like
quarks are required to be similar or above $M_{W'}$ to get an enhancement in the $WZ$ mode. 
Otherwise, such mixings would lead to new decay modes of $W'$ decaying to a vector-like quark and a 
SM quark with large branching fractions.}
\begin{align}
\begin{pmatrix} u_{i}^{\prime} \\  xu_{i}^{\prime} \end{pmatrix}~ = 
\begin{pmatrix} \cos\theta_{R}^{u_i} & \sin\theta_{R}^{u_i} \\ -\sin\theta_{R}^{u_i} & \cos\theta_{R}^{u_i}  \end{pmatrix}~ \begin{pmatrix} u_{i} \\  xu_{i} \end{pmatrix}
\end{align}
%%%
and the resulting doublets are given as
%%%
\begin{align}
Q_{i}^{R\prime} = \begin{pmatrix} u_{i}\cos\theta_{R}^{u_i}+xu_{i}\sin\theta_{R}^{u_i} \\  
d_{i} \end{pmatrix},   \hspace{1cm}
XQ_{i}^{R\prime} = \begin{pmatrix} -u_{i}\sin\theta_{R}^{u_i}+xu_{i}\cos\theta_{R}^{u_i} \\ xd_{i}\end{pmatrix}~.~\,
\end{align}
We do not consider the mixings in the 1st generation quarks which helps in keeping the production cross section 
for $W'$ unaffected, i.e., we will work with $\cos(\theta _{R}^{u})=1$. For illustration we choose 
$\cos(\theta_{R}^{t})=0.8$ here and vary $0.5<\cos(\theta_R^{c})<1$. Note that we could equally 
choose $\cos(\theta_{R}^{t})=1$ or vary it and we can still get a different but viable range for the parameter 
space. A similar mixing could 
also be chosen for the down quark sector which might also help in suppressing contributions to 
strangeness violating decays. 
We do not take up this issue here and focus only on the diboson excess.  
In Fig. \ref{fig:ccRswz-8TeV}, we show the allowed parameter space for the mixing 
angle $\cos(\theta_R^{c})$ which gives the desired range of 
$\sigma(pp \rightarrow W' )\times Br(W'\rightarrow WZ)$ taking values between
 5 and 10 fb. Note that the $\tan\phi$ values lie between 0.85 and 1.15. As $\tan\phi$ increases 
 the production cross section falls and therefore the rates decrease 
 as shown in the heat map in Fig.~\ref{fig:ccRswz-8TeV} where for $\tan\phi=1.15$ the diboson 
 rate approaches the lower end of 5 fb. Note that  for larger values of $\tan\phi$ one requires 
 smaller $\cos\theta_{R}^{c}$ to suppress the dijet branching and increase the $WZ$ branching fraction.
 Also a slightly larger $\tan\phi$ can be accommodated if both $\cos\theta^c_R$ and $\cos\theta^t_R$ 
 are allowed to vary together. 
 We also  show ${\sigma(pp \rightarrow W')\times Br(W'\rightarrow JJ)}$ for the allowed 
 parameter region of the scan in Fig.~\ref{fig:ccRsJJ-8TeV}.
With the run-II of LHC already collecting data at center of mass energy of 13 TeV, a good 
starting point would be confirm whether any excess observed in the run-I data in various final states 
or any bumps in invariant mass distributions pointing at physics beyond the SM were not mere 
fluctuations or misinterpretations of the data. This would also put counter checks on the new 
physics models that are proposed to explain the aforementioned hints of new, beyond SM physics.  
To check the validity of our model at the LHC run-II, we estimate the ($\sigma\times Br$) for $WZ$ and 
$JJ$ final states for the same range of parameters that could explain the diboson excess observed
at run-I. The corresponding results we obtain have been shown in 
Fig.~\ref{fig:ccRswz-13TeV} and Fig.~\ref{fig:ccRsJJ-13TeV} respectively. Quite clearly, 
the allowed model parameters give a dibson resonance at 2 TeV with a parton cross section for
WZ in the range 50-65 fb, while the dijet resonance rate should be less than 950 fb. The above numbers
can be easily confirmed to check our model predictions for the observed dibson anomaly at LHC run-II.
\begin{figure*}[t]
\centering
\begin{minipage}[b]{.45\textwidth}
\includegraphics[height=0.9\linewidth,width=1.1\linewidth]{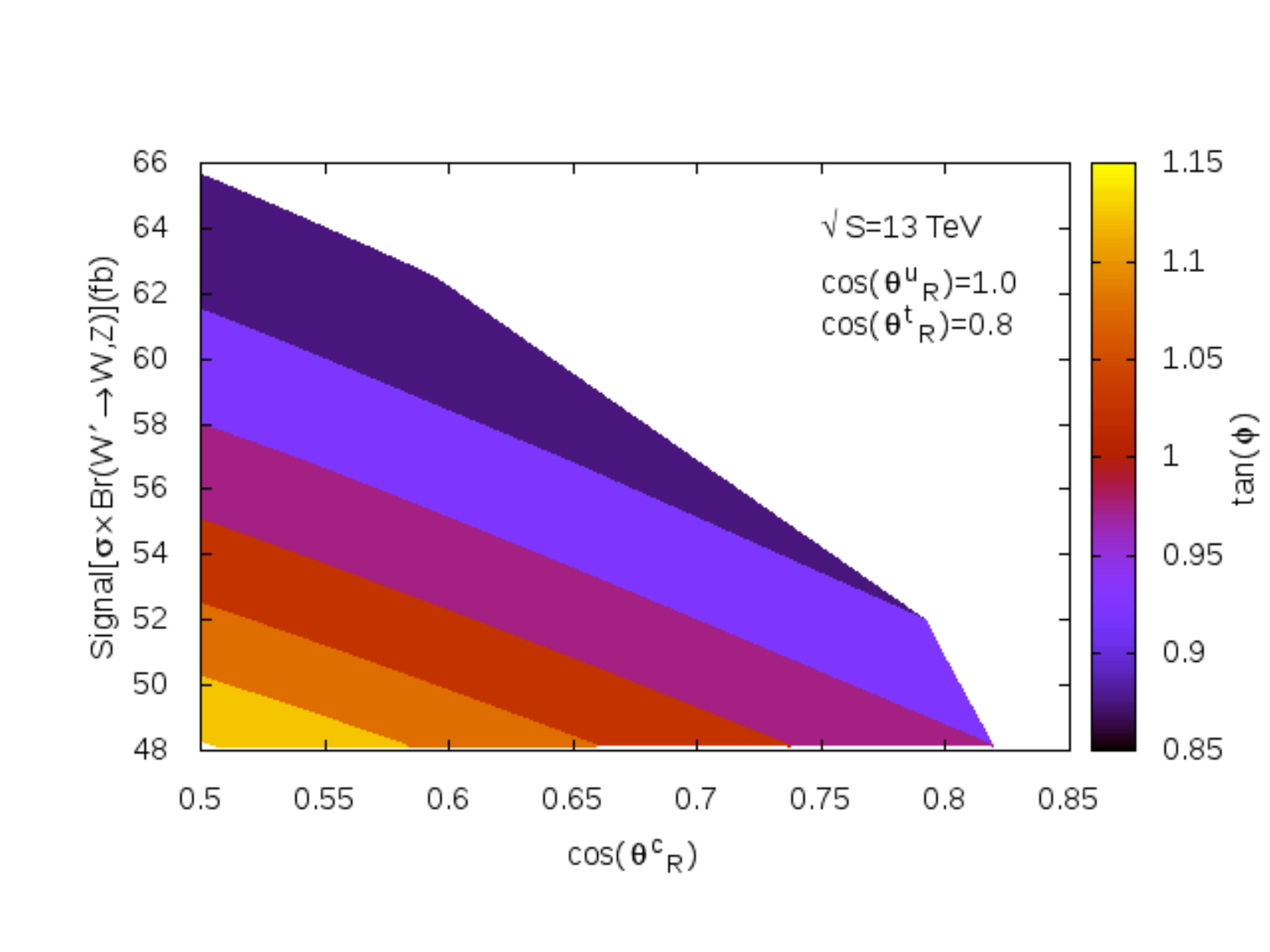}
 \caption{Signal for cross section times branching ratio for  $W^{\prime}\rightarrow WZ$ versus $\cos(\theta_{R}^{c})$ for different values of $\tan\phi$.} 
 \label{fig:ccRswz-13TeV}
\end{minipage}\qquad
\begin{minipage}[b]{.45\textwidth}
\includegraphics[height=0.9\linewidth,width=1.1\linewidth]{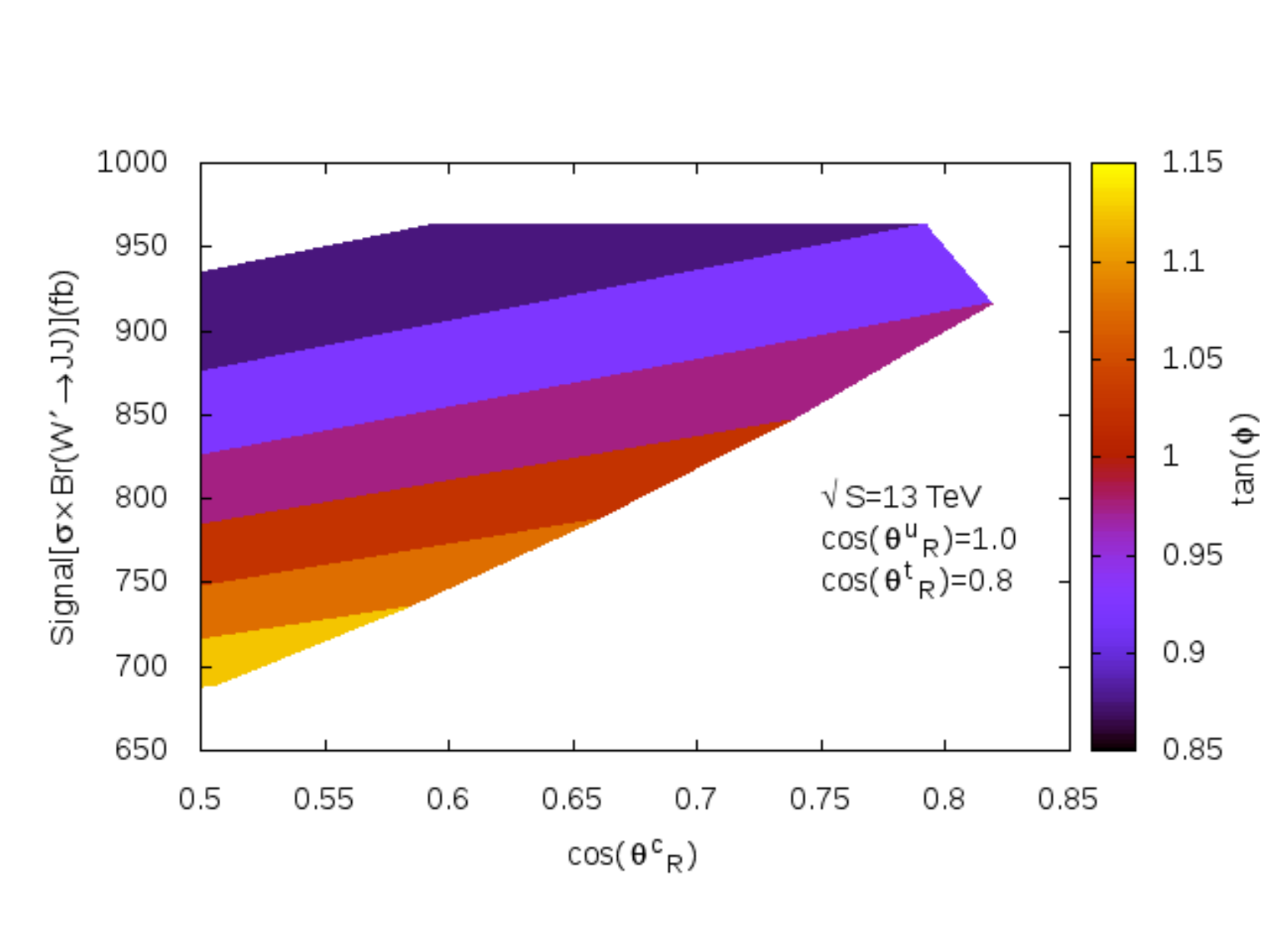}
 \caption{Signal for cross section times branching ratio for  $W^{\prime}\rightarrow jj$ versus $\cos(\theta_{R}^{c})$ for different values of $\tan\phi$.} 
 \label{fig:ccRsJJ-13TeV}
\end{minipage}
\end{figure*}
%%%%%

We now turn our attention to the other heavy gauge boson ($Z'$) in the theory. Note that we had assumed that
the mass of the heavy $Z'$ is above 2.5 TeV and chosen $\tan\phi$ accordingly. As the $Z'$ is 
an admixture of the $U(1)_X$ and the neutral components of $SU(2)_L$ and $SU(2)_R$ gauge bosons,
it cannot be completely leptophobic like the $W'$. The strong limits on a $Z'$ decaying into the leptonic final 
states put strong constraints on its mass.      
Therefore we need to check whether the $Z'$ in our model is  indeed required to be much heavier than the
$W'$ in satisfying all the existing constraints. We do the analysis by keeping $M_{W^\prime} = 2$ TeV. 
As the $Z'$ mass is dependent on the choice of $\tan\phi$ as shown in Fig.~\ref{fig:Zpmass}, 
we vary $\tan \phi$ in the range $0.05-3$. For this range of $\tan \phi$, $M_{Z'}$ varies from 2 TeV 
to 6.3 TeV following Eq.~(\ref{mzp_bp1}). However, we have already noted that for $W'$ 
to have a narrow width, $\tan\phi \geq 0.31$. Thus, the lower choices of $\tan\phi$ are for 
illustration purposes only and to also highlight the features when the mass is close to the $W'$ mass. 
As the interactions of $Z'$ with fermions as well as gauge bosons are independent of 
$\beta$ (See Eqs.~(\ref{eqn:Zprime-ffbar})-(\ref{FM-Z}).), the value of $\tan \beta$ does not play a significant 
role in the $Z'$ phenomenology.  Note that the $\tan\phi$ value 
gives an idea on the width of the $W'$ as well as for the $Z'$. The value of $\tan\phi$ 
%%%%%%%%
\begin{figure*}[t]
\centering
\begin{minipage}[b]{.45\textwidth}
\includegraphics[height=0.8\linewidth,width=1.0\linewidth]{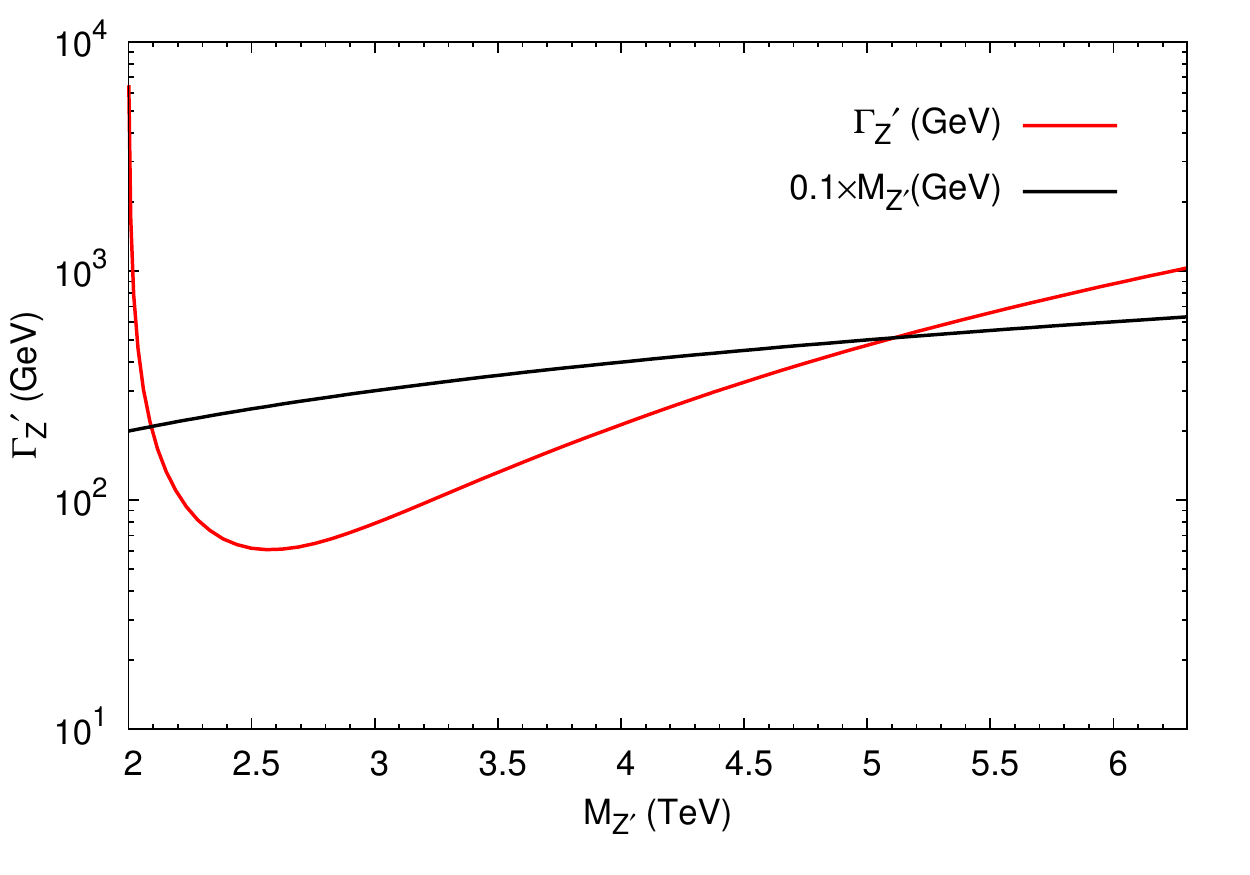}
  \caption{Width of Z$^\prime$ versus $M_{Z^\prime}$}
 \label{fig:Zprime-Width-MZp}
\end{minipage}\qquad
\begin{minipage}[b]{.45\textwidth}
\includegraphics[height=0.8\linewidth,width=1.0\linewidth]{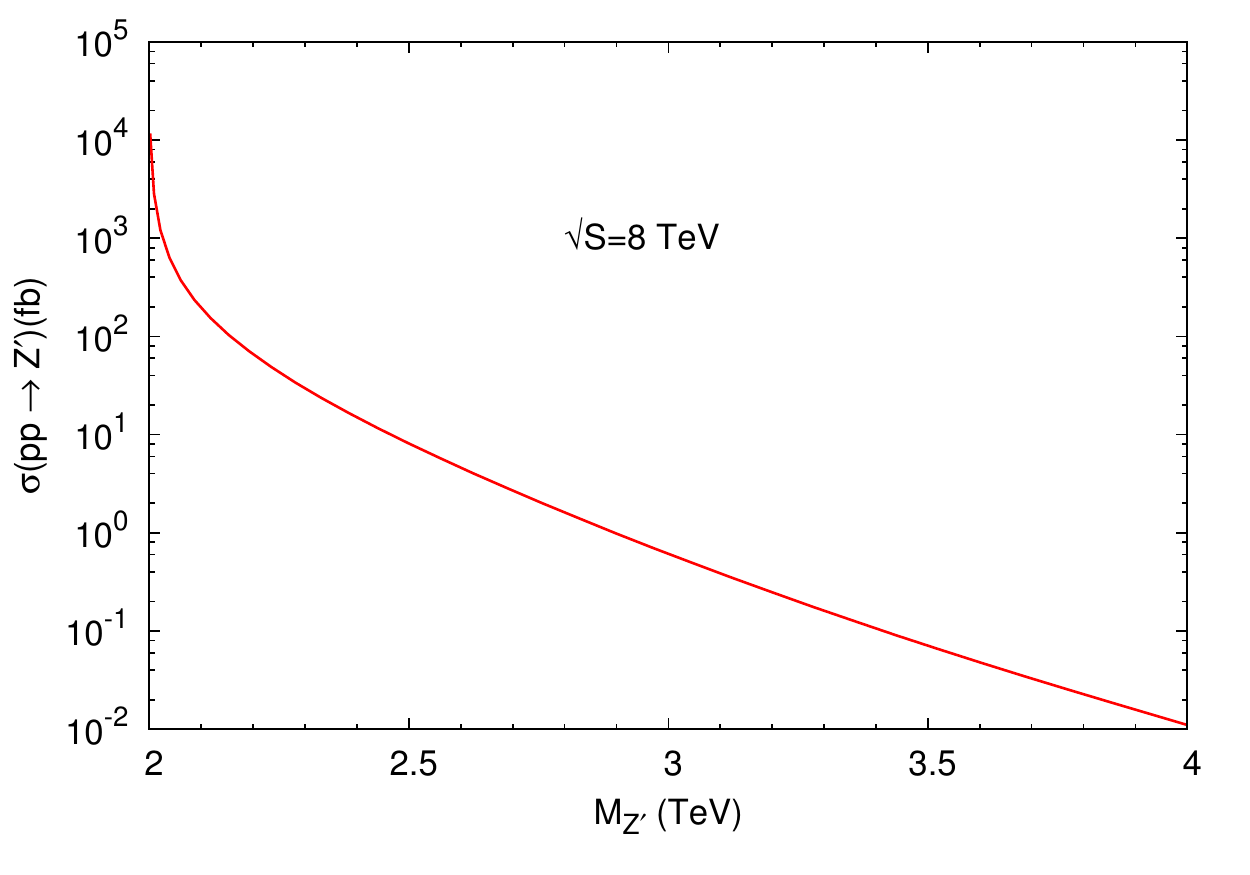}
\caption{Production cross section of Z$^\prime$.} 
 \label{fig:Zprime-prod-8TeV-MZp}
\end{minipage}
\end{figure*}
%%%%%
also determines how large the $SU(2)_R$ and $U(1)_X$ gauge couplings are and therefore 
would give us the relative characteristics of its leptophobic nature. To obtain the range 
over which the $Z'$ width is less than $0.1 \times M_{Z'}$ we plot the total width of $Z'$ in  
Fig.~\ref{fig:Zprime-Width-MZp} for different values of $M_{Z'}$.  
It is clear from Fig.~\ref{fig:Zprime-Width-MZp} that narrow width approximation for $Z'$ is valid for 
the mass range of 2.1 TeV to 5.1 TeV which implies $\tan\phi$ to be in the range $\sim 0.3 - 2.35$. 
Note that in the limit $x \gg 1$
\begin{align*}
  M_{Z'} \simeq M_{W'}/\cos\phi~,
\end{align*}
which implies that the $Z'$ is always heavier than the $W'$. Thus, a 2 TeV $Z'$ would lead to 
a much broader resonance for the $W'$ and itself.  
 We plot the on-shell production cross section of the $Z'$ at LHC with $\sqrt{s}=8$ 
 TeV in Fig.~\ref{fig:Zprime-prod-8TeV-MZp}.  In the range where the $Z'$ has 
 a small width to ensure narrow width approximation, we can estimate the rates 
 for different final states by multiplying with the corresponding branching 
 fractions. Note that with the narrow width criterion, the  $M_{Z'}$ can be as light as 2.1 TeV if it 
 satisfies the current limits set on the different channels. In order to satisfy the diboson excess, 
 the $Z'$ contributions therefore could also contribute to the excess. However, the large production 
 rates for the dijet channel through the $W'$ production could restrict such values for $\tan\phi$. 
 We then might need to include the mixing of the vector-like quarks with the first generation 
 quarks to suppress the dijet rates.  
 
 %%%%%%%%
\begin{figure*}[t]
\centering
\begin{minipage}[b]{.45\textwidth}
\includegraphics[height=0.8\linewidth,width=1.0\linewidth]{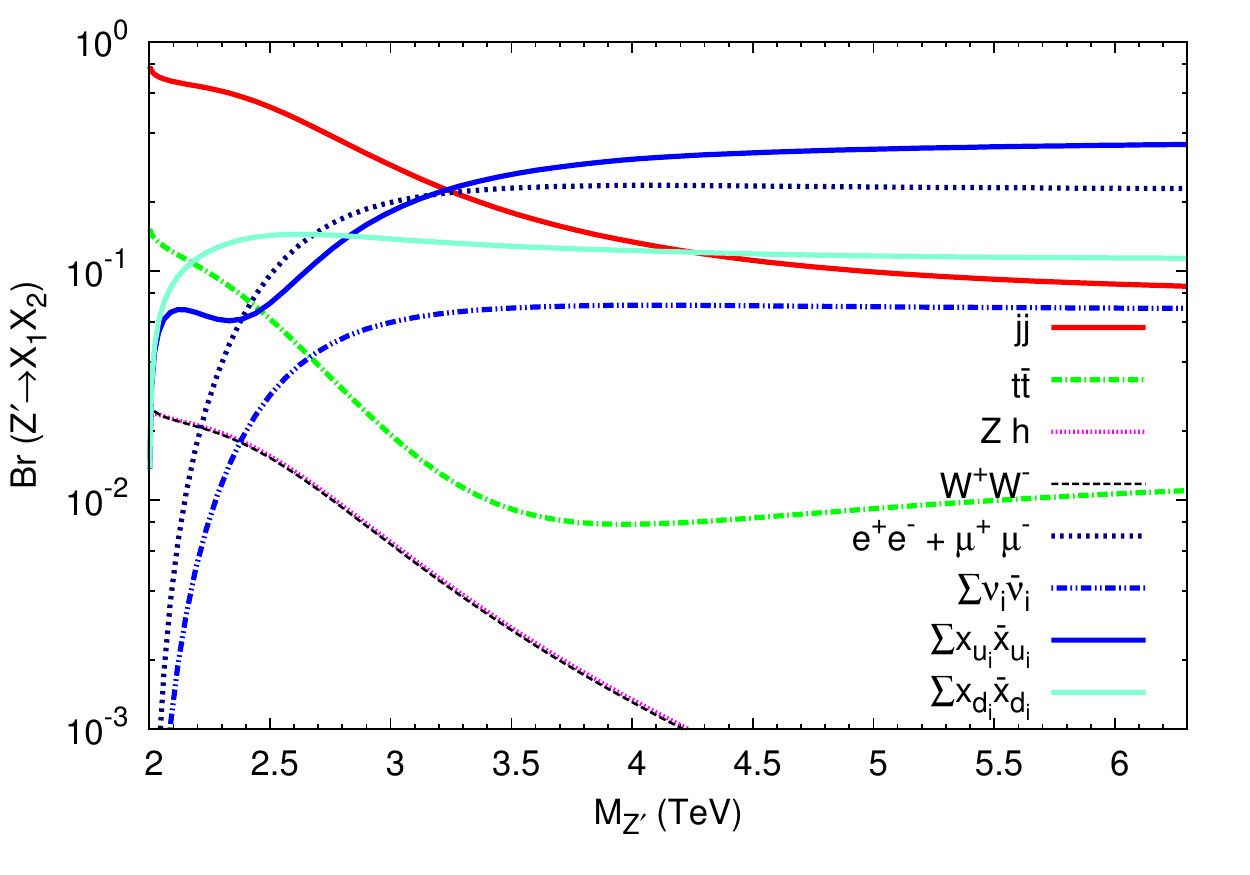}
  \caption{Branching ratio for different decay modes for Z$^\prime$ versus $M_{Z^\prime}$.} 
 \label{fig:Zprime-Br-MZp}
\end{minipage}\qquad
\begin{minipage}[b]{.45\textwidth}
\includegraphics[height=0.8\linewidth,width=1.0\linewidth]{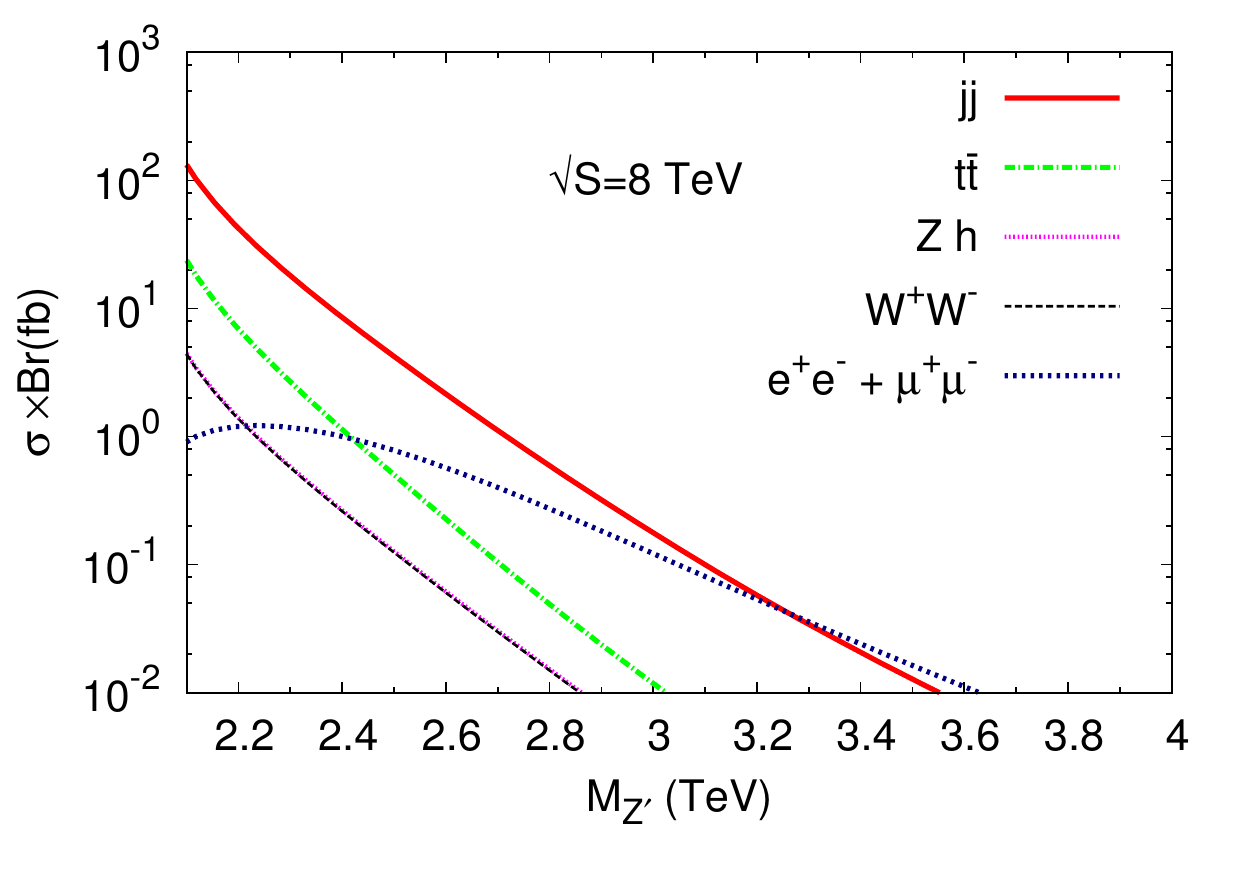}
\caption{Cross section times branching ratio for Z$^\prime$ versus $M_{Z^\prime}$.} 
 \label{fig:Zprime-XS-Br-8TeV-MZp}
\end{minipage}
\end{figure*}
%%%%%
 To check the signal strength for different channels we 
 have calculated the branching fractions of the $Z'$ decay which we show in
Fig.~\ref{fig:Zprime-Br-MZp}. As the interaction strength of $Z^\prime$ to leptons 
increases with $\tan \phi$ (see Eqs.~(\ref{eqn:Zprime-ffbar})-(\ref{eqn:Zprime-LR})), this leads to 
the increase in  
branching ratio for $Z^\prime$ decaying to leptons. This is because the $U(1)_X$ 
gauge coupling becomes larger for larger values of $\tan\phi$ thus making the $Z'$ 
less leptophobic. However with increasing $\tan\phi$ the $Z'$ mass also 
increases and therefore the dilepton channel will not be strongly constrained 
for such a heavy $Z'$ by the run-I data at LHC.
In Fig. \ref{fig:Zprime-XS-Br-8TeV-MZp} we have plotted $(\sigma \times BR)$ for different final states.
The upper limit for dijet resonance given in the CMS dijet analysis~\cite{Khachatryan:2015sja}
is satisfied for 
$M_{Z^\prime}>$ 2.3 TeV. However, allowing mixing between the vector-like quarks 
with the SM quarks will again dilute the dijet rates and allow a slightly lighter $Z'$.  
This mixing would further increase the branching fraction of the $Z'$ decaying 
leptonically and therefore  beyond $M_{Z'} > 2.4 $ TeV, it would provide the 
best mode of discovery at the current run-II of LHC~\footnote{With the increase in leptonic branching
for a heavier $Z'$, one can in principle accommodate the 2.9 TeV anomalous 
resonant event reported in the early data of run-II by CMS in the leptonic final state~\cite{CMS-3TeV} in our model.}.
In addition we find that a dominant mode of decay for the $Z'$ is to a pair of vector-like quarks. Now, 
the exact rates would depend on the mass of these exotic quarks as well as the mixings they posess
with SM like quarks.  But the most interesting aspect would be the resonant production of such 
colored exotics through a $Z'$ leading to enhanced production rates for a pair as well as 
single production modes (when mixing with SM quarks is subtantial) which could give new 
signals at the run-II of LHC and its future runs.
%%%%%%%%%%%%%%%%%%%%%%%%%%%%%%%%%%%%%%%%%%%%%%%%%%%%

%%%%%%%%%%%%%%%%%%%%%%%%%%%%%%%%%%%%%%%%%%%%%%%%%%%%

\section{Conclusion}

We  employed the left-right models to explain the ATLAS diboson excess.
To escape the tight constraints from lepton plus missing energy searches, we required 
the $SU(2)_R$ gauge symmetry to be leptophobic. However, in the previously considered 
models, anomaly cancellations have not been ensured. Therefore, we for the first time propose 
an anomaly free leptophobic left-right model with gauge symmetry 
$SU(3)_C\times SU(2)_L \times SU(2)_R \times U(1)_{X}$, where the SM leptons are singlets 
under $SU(2)_R$. To cancel the gauge anomalies, we introduced the extra vector-like quarks. 
Since the $Z'$ gauge boson cannot be leptophobic, we assume its mass to be around or 
above 2.5 TeV and then the constraint on dilepton final state can be avoided.
In addition, we found that the $W'\to WZ$ channel cannot explain the ATLAS diboson excess
if we included the
constraint on $W'\to jj$ decay mode. Interestingly, we solved this problem by considering the
mixings between the SM quarks and vector-like quarks.
We showed explicitly that the ATLAS diboson excess can be explained in the viable parameter space
of our model, which is consistent with all the current experimental constraints. In addition, we have 
also given predictions for the dijet and $WZ$ channel at the current run of LHC with $\sqrt{s}=13$ TeV
and discussed the ensuing phenomenology of $Z'$ for the viable parameter space of our model. We
also propose new signals for the vector-like quarks in our model which can be studied at the high 
energy run of the LHC.  

%%%%%%%%%%%%%%%%%%%%%%%%%%%%%%%%%%%%%%%%%%%%%%%%%%%

\begin{acknowledgments}
This research was supported in part by the Department of Atomic Energy, Government of India, for the Regional Centre for Accelerator-based Particle Physics (RECAPP), 
Harish-Chandra Research Institute (KD, SKR), by the Natural Science Foundation of China under 
grant numbers 11135003, 11275246, and 11475238 (TL) and by the US Department of Energy Grant 
Number DE-SC0010108 (SN).
\end{acknowledgments}

\end{document}